# Indoor Localization of Smartphones Thanks to Zero-Energy-Devices Beacons

Shanglin Yang*†o, Yohann Bénédic*‡, Dinh-Thuy Phan-Huy*, Jean-Marie Gorceo, Guillaume Villemaudo

*Orange Innnovation/Networks, †Châtillon, ‡Belfort, France  {name.surname}@orange.com

o INSA Lyon, Inria, CITI, EA3720, Villeurbanne, France {name.surname}@insa-lyon.fr

*Abstract*—In this paper, we present a new ultra-low power method of indoor localization of smartphones (SM) based on zero-energy-devices (ZEDs) beacons instead of active wireless beacons. Each ZED is equipped with a unique identification number coded into a bit-sequence, and its precise position on the map is recorded. An SM inside the building is assumed to have access to the map of ZEDs. The ZED backscatters ambient waves from base stations (BSs) of the cellular network. The SM detects the ZED message in the variations of the received ambient signal from the BS. We accurately simulate the ambient waves from a BS of Orange 4G commercial network, inside an existing large building covered with ZED beacons, thanks to a ray-tracing-based propagation simulation tool. Our first performance evaluation study shows that the proposed localization system enables us to determine in which room a SM is located, in a realistic and challenging propagation scenario.

## I. INTRODUCTION

Indoor self-localization of smartphones (SM) can still be improved in terms of cost and energy efficiency. As Global Positioning System (GPS) based localization only works properly outdoor [1], many alternatives have been studied and even commercialized based on terrestrial wireless networks. Indeed, wireless network nodes with known locations can be used to locate the smartphone (SM), either through the analysis of downlink signals by the SM or through the analysis of uplink signals by the network side. In practice, 5th generation (5G) cellular networks macro base stations (BSs) [2], 5G small cells [3], wi-fi access points [4] or Blue Tooth Low Energy (BLE) [5][6] have been successfully tested, with several meters of accuracy. However, they require a strong density of nodes and/or highly complex radio finger-printing schemes. Unfortunately, as all these nodes need power supply, and thus, their deployment cannot be massive.

Recently, the first European Flagship Project Hexa-X I on the future sixth generation (6G) of networks has set as a goal, to provide digital services (including localization) sustainably [7] and introduced a new type of device called Crowd-Detectable zero-energy-devices (CD-ZED) [8]. A CD-ZED, or simply ZED, is a self-powered device that harvests ambient energy to power itself and that backscatters (reflects) ambient waves from the base stations (BSs) of the cellular network to communicate with a SM [9]. In practice, the ZED switches between a transparent state and a reflective state to transmit a bit '0' or a bit '1'. A SM closed by, detects the ZED message by analyzing variations in the received signal issued from the BS. Very recently, several material proofs-of-concept and experiments have been successfully conducted with 4th Generation (4G) base stations as sources of ambient waves (as 4G is a mature ambient network), and pilot-based detection at the SM side [10-12]. Furthermore, [9] presents a ZED prototype that powers itself with solar and indoor light energy, stores some energy in a small battery, and keeps transmitting data night and day. Finally, such types of devices are currently being studied at the 3rd generation partnership project (3GPP) for potential introduction in future mobile networks standards, under the umbrella concept of "Ambient Internet-of-Things" (A-IoT) [13].

In this paper, we propose for the first time a new Indoor localization method relying on the ZED technology proposed in [9]. Such method is expected to be more sustainable. In our study, 4G is used as an ambient source of signal, as it is the most mature network today. However, similar studies can be done with 5G or beyond standards. The paper is organized as follows: Section II presents our ZED-to-SM communication model, Section III presents our proposed ZED-based localization scheme, Section IV presents our simulation-based performance evaluation method, Section V presents results and Section VI concludes this paper.

## II. ZED-TO-SM COMMUNICATION MODEL

Sub-section A provides an overview of the ZED-to-SM communication system. Sub-section B focuses on the propagation model, sub-section C details the transmission, detection and Signal-to-noise ratio (SNR) estimation schemes, and sub-section D provides an estimation of the Bit Error Probability (BEP) and the SNR.

### A. Overview

We consider the communication between a ZED and a SM illuminated by an ambient 4G Base Station (BS) transmitting an OFDM waveform over $N^{RB}$ resource blocks (RBs), every sub-frame of TTI (Transmission Time Interval) duration [14]. The BS sends K pilot symbols in each RB, periodically, every subframe, on standardized locations in time and frequency (i.e. on pre-defined sub-carriers and OFDM symbols, inside the RB). Without loss of generality, we assume that all pilot symbols have the same constant magnitude $\sqrt{P^u}$, where $P^u$ is the pilot transmit power spectral density.

As in [9], the ZED transmits bit $b = 0$, the ZED by being in a 'transparent mode' where it is transparent to ambient waves from the BS. As in [9], the ZED transmits bit $b = 1$ by being in a 'reflecting mode' where it backscatters (reflects in all directions) ambient waves from the BS. In our current study, the ZED transmits one bit during one period of $N^{TTI}$ consecutive subframes. The bit period duration is therefore $T^b = N^{TTI}TTI$ and the bit rate (without loss due to the synchronization phase) is $R^b = \frac{1}{T^b}$. Finally, the ZED sends a data sequence of $N^b$ successive bits.

The SM monitors the pilots from the BS, estimates the downlink propagation channel, and searches for the ZED message in the channel estimates. More precisely, we define $\mathbf{y} \in \mathbb{C}^{N \times 1}$ as the vector of received pilot symbols during one bit period $T^b$ as follows : $\mathbf{y}_{(n^{TTI}-1)N^{TTI}N^{RB}K+(n^{RB}-1)N^{RB}K+k}$

is the $k^{th}$ received pilot symbol in RB $n^{RB}$ and during sub-frame $n^{TTI}$, with $1 \leq n^{TTI} \leq N^{TTI}$ and $1 \leq n^{RB} \leq N^{RB}$. The total number of received pilot symbols per bit is therefore $N = N^{RB} N^{TTI} K$. With these definitions, $\mathbf{y} \in \mathbb{C}^{N \times 1}$ can be expressed as follows:

$$\mathbf{y} = \mathbf{x} + \mathbf{w}, \qquad (1)$$

where $\mathbf{x} \in \mathbb{C}^{N \times 1}$ is the useful part of the received pilot symbols and $\mathbf{w} \in \mathbb{C}^{N \times 1}$ is the complex additive white gaussian noise (AWGN) with power spectral density $N_0$. We define $\mathbf{h} \in \mathbb{C}^N$ as the vector of propagation channel coefficients for the N pilot symbols. We include the antenna gains of the BS and the SM inside $\mathbf{h}$. With these notations and assumptions, the expression of $\mathbf{x}$ is given by:

$$\mathbf{x} = \sqrt{P^u}\mathbf{h}. \qquad (2)$$

Depending on the state of the ZED, $\mathbf{h}$ can take two values: $\mathbf{h} = \mathbf{\Gamma}$ when ZED is in transparent mode and sending bit '0'; $\mathbf{h} = \mathbf{G}$ when ZED is in reflecting mode and sending bit '1'; where $\mathbf{\Gamma}, \mathbf{G} \in \mathbb{C}^{N \times 1}$ are computed as explained in Sub-Section B.

Depending on the state of the ZED, $\mathbf{x}$ can also take two values: $\mathbf{x}^T = \sqrt{P^u}\mathbf{\Gamma} \in \mathbb{C}^{N \times 1}$ when ZED is in transparent mode and $b = 0$; $\mathbf{x}^R = \sqrt{P^u}\mathbf{G} \in \mathbb{C}^{N \times 1}$ when ZED is in reflecting mode and $b = 1$. Upon reception of $\mathbf{y}$, the SM estimates $\mathbf{x}$ and $b$ as described in Sub-Section C.

### B. Propagation channel model

In the propagation channel between the BS and the SM, we must consider two propagation paths, as depicted in Fig. 1: the direct path from the BS to the SM, when the ZED is in transparent mode; the backscattered path from the BS to the ZED, then from the ZED to the SM, when the ZED is in reflecting mode. A ZED reflects a small amount and a large amount of waves, in transparent mode, and in reflecting mode, respectively [9]. In this paper, we assume an ideal ZED that reflects nothing in transparent mode and that reflects everything in reflecting mode. We also assume that the BS and the UE have one antenna. Under these assumptions, the expression of $\mathbf{h}$ is given by: $\mathbf{h} = \mathbf{\Gamma}$ when ZED is in transparent mode; $\mathbf{h} = \mathbf{G} = \mathbf{\Gamma} + \mathbf{\Phi}.\mathbf{\Lambda}$ when ZED is in reflecting mode, where $(.)$ is the Hadamard matricial product, $\mathbf{\Gamma}, \mathbf{\Phi}, \mathbf{\Lambda} \in \mathbb{C}^{N \times 1}$ are the vectors of the channel coefficients (for the N pilots and including antenna gains), for the direct BS-to-SM path, the BS-to-ZED path and the ZED-to-SM path, respectively.

### C. Transmission, reception, and SNR estimation

We propose a communication system between the ZED and SM composed of two phases: a learning and synchronization phase; and a data communication phase. We assume that the channel is constant during the learning and synchronization phase, and the whole communication phase.

During the learning and synchronization phase, the ZED sends a training bit-sequence $\mathbf{S}_0 = [0 \ \ldots \ 0]$ followed by a training bit-sequence $\mathbf{S}_1 = [1 \ \ldots \ 1]$ of predefined size L known by the SM. The SM stores and continuously updates a set $\mathbf{Y}$ of 2L successive values of the received symbol $\mathbf{y}$, $\mathbf{Y} = [\mathbf{y}^{(1)} \ \ldots \ \mathbf{y}^{(2L)}]$ and computes $\mu = |\sum_{l=1}^{L} \mathbf{y}^{(l+L)} - \mathbf{y}^{(l)}| / \sum_{l=1}^{L} |\mathbf{y}^{(l+L)} - \mathbf{y}^{(l)}|$. The SM uses the criterium $\mu > 0.9$ to detect $\mathbf{S}_0 \ \mathbf{S}_1$. Then, the SM estimates $\mathbf{x}^T$ and $\mathbf{x}^R$ as follows:

$\hat{\mathbf{x}}^T = \frac{1}{L}\sum_{l=1}^{L} \mathbf{y}^{(l)}$ and $\hat{\mathbf{x}}^R = \frac{1}{L}\sum_{l=1}^{L} \mathbf{y}^{(l+L)}$. From now on, we assume that the SM is perfectly bit-synchronized with the ZED and that $\mathbf{x}^T$ and $\mathbf{x}^R$ are perfectly estimated (i.e. $\hat{\mathbf{x}}^T = \mathbf{x}^T$ and $\hat{\mathbf{x}}^R = \mathbf{x}^R$).

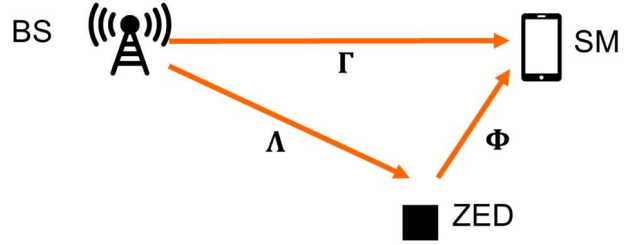

Fig. 1. Channel model.

Then, during the data communication phase, for each $T^b$ period, the SM processes the received $\mathbf{y}$ values to detect the received bit, until it has received the $N^b$ bits of the full data sequence. To do so, the SM uses a coherent detector [16] as follows. The SM first computes:

$$\mathbf{m} = \frac{\mathbf{x}^T + \mathbf{x}^R}{2} \in \mathbb{C}^{N \times 1} \text{ and } \mathbf{u} = \frac{\mathbf{x}^T - \mathbf{x}^R}{\|\mathbf{x}^T - \mathbf{x}^R\|} \in \mathbb{C}^{N \times 1}.$$

where $\|.\|$ is the Frobenius norm. Then, for each transmitted bit b, the SM receives $\mathbf{y}$ and computes $\mathbf{v} \in \mathbb{C}^{N \times 1}$ and $r \in \mathbb{R}$ as follows:

$$\mathbf{v} = \mathbf{y} + \mathbf{m} \qquad (3)$$

$$r = \mathbf{u}^H.\mathbf{v} \qquad (4)$$

where $\mathbf{u}^H$ is the Hermitian of $\mathbf{u}$. Finally, the estimates $\hat{\mathbf{x}}$ and $\hat{b}$ are determined by the SM as follows:

- if $r > 0$, then $\hat{\mathbf{x}} = \mathbf{x}^T$ and $\hat{b} = 0$;
- otherwise, $\hat{\mathbf{x}} = \mathbf{x}^R$ and $\hat{b} = 1$.

We introduce $C = \frac{1}{N}\|\mathbf{x}^T - \mathbf{x}^R\|^2 = \frac{P^u}{2}\|\mathbf{F} - \mathbf{G}\|^2 = \frac{P^u}{N}\|\mathbf{\Phi}.\mathbf{\Lambda}\|^2$. $C$ is in fact the received power from the ZED alone. We, hence, define the ZED SNR as follows:

$$\frac{C}{N_0} = \frac{1}{N_0}\frac{\|\mathbf{x}^T - \mathbf{x}^R\|^2}{N} = \frac{P^u}{N_0}\frac{\|\mathbf{\Phi}\mathbf{\Lambda}\|^2}{N}. \qquad (5)$$

We assume that the SM estimates perfectly the SNR based on based on the perfect estimates of $\mathbf{x}^T$, $\mathbf{x}^R$ and $N_0$.

### D. BEP

The BEP of the coherent detector described in Sub-Section C, is given by [15]:

$$\text{BEP} = Q\left(\frac{\|\mathbf{x}^T - \mathbf{x}^R\|}{2\sqrt{2N_0}}\right), \qquad (6)$$

where $Q(.)$ is the Q-function. By using (5), we obtain:

$$BEP = Q\left(\frac{1}{2\sqrt{2}}\sqrt{N\frac{C}{N_0}}\right), \qquad (7)$$

As expected, the BEP increases with the SNR. We also remark that, using (5), the BEP is likely to be better when the received power $\frac{P^u}{N}\|\mathbf{\Lambda}\|^2$ and the ZED-to-SM propagation path power $\|\mathbf{\Phi}\|^2$ are strong. Finally, as $N = N^{RB}N^{TTI}K$, an increase in $N^{TTI}$ is expected to provide a redundancy gain.

## III. ZED BASED LOCALIZATION SYSTEM MODEL

In this section, we detail our proposed ZED-based system for the localization of the SM inside a building.

In Fig. 2, we illustrate the ZED-based localization scheme. $N_{ZED}$ ZEDs are deployed inside a building. Each ZED is equipped with a unique identification number coded into a bit-sequence, and its precise position on the map is recorded. An SM inside the building is assumed to have the map of ZEDs. Additionally, the building is in the coverage of a BS. ZEDs are assumed to transmit their identification bit-sequence periodically, but at different times, to avoid collisions. The SM monitors the received pilots from the BS, detects, successively, the messages of ZEDs nearby, and computes their respective SNRs, as described in Section II. Then, the SM determines the ZED with the strongest SNR and considers it as the closest one. Finally, the SM locates itself on the map at the same location as the selected ZED. Fig. 2-a) illustrates the waves transmitted by the BS alone, when the ZEDs near the SM are inactive. Fig. 2-b) illustrates the case where the SM detects ZED#2 with a weak SNR. Fig. 2-c) illustrates the case where the SM detects ZED#7 with the strongest SNR and deduces its location on the map.

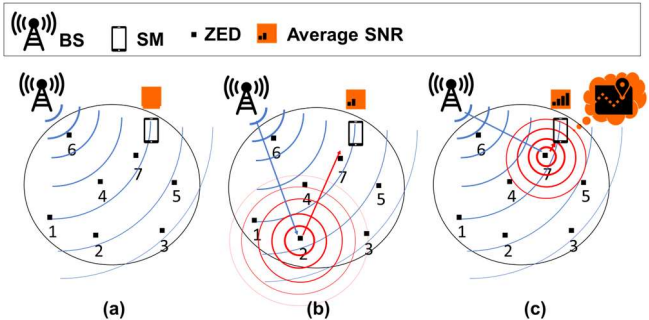

Fig. 2. Localization scheme principle : (a) SM monitors the received signal from the BS and detects no ZED ; (b) SM detects ZED#2 with a low SNR ; (c) SM detects ZED#7 with the strongest SNR, thus as the closest one, and deduces its location on the map.

## IV. PERFORMANCE EVALUATION METHODOLOGY

In this Section, we present our simulation-based method, to assess the ability of the SM to locate itself whatever its position inside the building, thanks to our proposed localization system (described in Section III). Sub-Section A describes the deployment scenario and simulation assumptions, and Sub-Section B focuses on the ray-tracing-based channel model.

### A. Deployment Scenario and Simulation Assumptions

We consider a real existing building under the coverage of a real BS from Orange 4G commercial network [17], both illustrated in Fig. 3. The BS has three azimuth orientations: 50°, 180° and 300° and transmits in the LTE band 852MHz-862MHz. Table I lists all the simulation parameters regarding the 4G BS. Other parameters (than those already defined in Section II and III) are defined hereafter. $P^{bs}$ is the BS transmit power, BW is the system bandwidth, $N^f$ is the SM noise figure, $N_{th}$ is the thermal noise density and $f_0$ is the central carrier frequency. The power spectral density $N_0$ is derived from BW, $N^f$ and $N_{th}$ and the power spectral density $P^u$ is derived from $P^{bs}$ and BW.

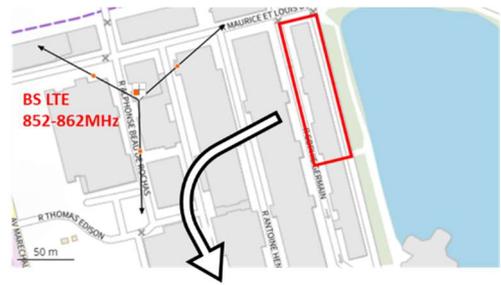

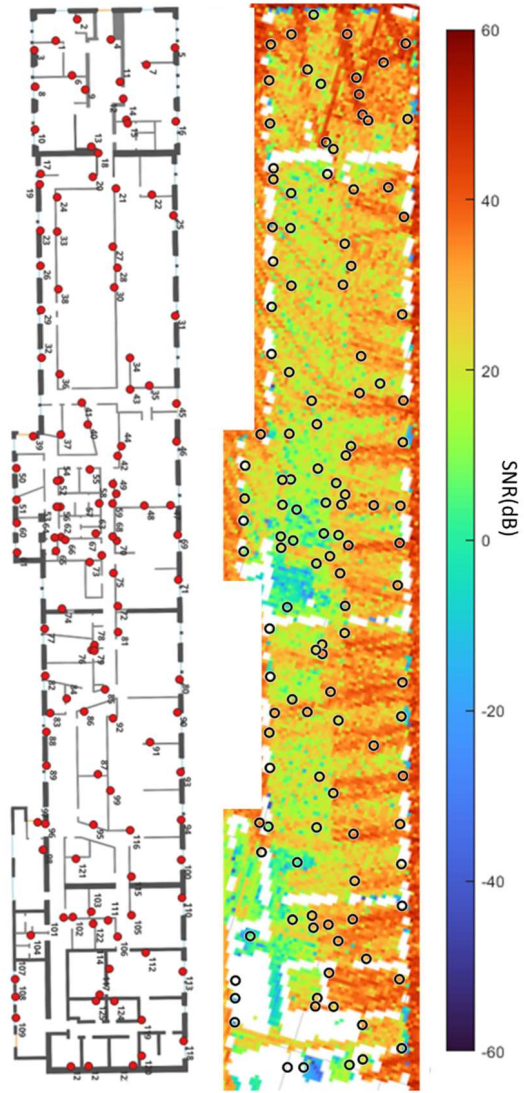

Fig. 3. Commercial 4G BS and building location [17], map of ZEDs deployed in the building, BS coverage inside the building

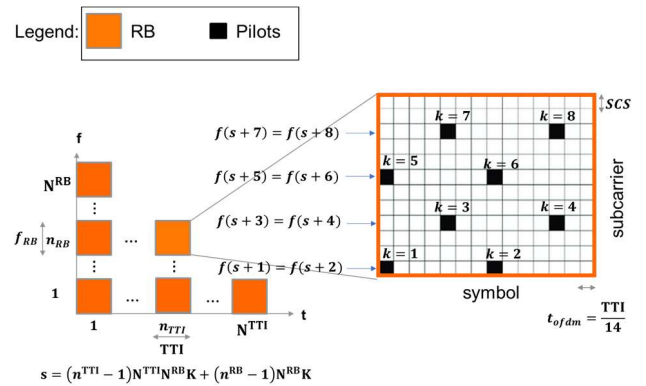

Fig. 4. Position of pilots in a RB and a TTI

$s = (n^{TTI} - 1)N^{TTI}N^{RB}K + (n^{RB} - 1)N^{RB}K$

The frequencies $f(n)$ of pilot symbol $n$, are calculated based on the 4G pilot standardized Time-Frequency pattern [14] illustrated by Fig. 4, where $f_{RB}$ is the frequency bandwidth of a RB, SCS is the subcarrier spacing, $t_{ofdm}$ is the duration of an OFDM symbol (with values provided in Table I). For each RB there is a total of K = 8 pilot symbols. Various numbers of TTIs $N_{TTI}$ are considered.

The deployment of ZEDs on the map is done manually, via the software QGIS [16]. $N_{ZED}$ = 127 ZEDs in total, are deployed in each room of the building, at positions with best illumination from the BS, as illustrated by Fig. 3.

TABLE I. PARAMETERS

| Parameters | $N_{RB}$ | $N_{TTI}$ | K | $t_{ofdm}$ |
|---|---|---|---|---|
| Value | 50 | 1, 2, 3, 6 | 8 | TTI/14 |
| Parameters | $f_{RB}$ | SCS | TTI | $N^f$ |
| Value | 180kHz | 15kHz | 1ms | 9dB |
| Parameters | BW | $p^{bs}$ | $N_{th}$ | |
| Value | 9MHz | 46dBm | -174dBm per Hz | |
| Parameters | $f_0$ | $N_{ZED}$ | BEP threshold | |
| Value | 857MHz | 127 | 0.01 | |

To assess the performance of the localization scheme, we divide the map into pixels of $40cm \times 40cm$. For each pixel, we first position the SM on the pixel, we then calculate the SNRs and the BEPs for each ZED, using the model of the ZED-to-SM communication described in section II. Finally, we find the closest ZED for this specific SM using the model of the localization system described in Section III. Additionally, though the ZED is not an active transmitter, it does have a coverage area (CA). This is the area where the BEP is better than the threshold parameter in Table I.

### A. Ray-tracing based channel model [18]

In order to evaluate the performance of our proposed scheme with a realistic propagation channel model, we use the Orange internal ray-tracing simulation tool STARLIGHT[18] to generate the complex channel gain $\alpha_p \in \mathbb{C}$ of each propagation path (ray) $p$ of the multipath channel, at the central carrier frequency $f_0$. $\alpha_p$ already includes the BS antenna gain and the effect of the antenna diagram, for LTE antenna port 0 [14], and the SM antenna gain (which is assumed to be 0dBi). Then, the complex channel gains $\mathbf{\Gamma}^n, \mathbf{G}^n \in \mathbb{C}$ are deduced as follows:

$$\mathbf{\Gamma}^n = \sum_{p=1}^{N_{path}^\Gamma} \alpha_p e^{-j2\pi(f(p)-f_0)\tau_p^\Gamma}, \quad (8)$$

$$\mathbf{G}^n = \mathbf{\Gamma}^n + \mathbf{\Phi}^n \mathbf{\Lambda}^n = \sum_{p=1}^{N_{path}^\Gamma} \alpha_p e^{-j2\pi(f(p)-f_0)\tau_p^\Phi} + \left(\sum_{p=1}^{N_{path}^\Phi} \beta_p e^{-j2\ (f(p)-f_0)\tau_p^\Phi}\right) \times \left(\sum_{p=1}^{N_{path}^\Lambda} \delta_p e^{-j2\pi(f(p)-f_0)\tau_p^\Lambda}\right), \quad (9)$$

where $N_{path}^\Gamma, N_{path}^\Phi, N_{path}^\Lambda$ are the total numbers of paths of the three $\mathbf{\Gamma}^n, \mathbf{\Phi}^n$ and $\mathbf{\Lambda}^n$ channels, $f(p)$ is the frequency of the sub-carrier on which is mapped the nth pilot used to send a bit, and $\tau_p^\Gamma, \tau_p^\Phi, \tau_p^\Lambda$ are the delays of the $p_{th}$ ray for the three channels, $\alpha_p, \beta_p, \delta_p \in \mathbb{C}$ are the complex channel gains for the three channels.

## V. SIMULATION RESULTS

In Fig. 5, we present maps of the closest detected ZED by the SM as a function of the SM's true position. These results are obtained for cases where the BEP is less than 0.01, and we compare the scenarios of $N_{TTI}$=1, 2, 3 and 6. Each black circle indicates the presence of a ZED. The color in each circle represents the ZED's identification, and the area covered by this color corresponds to the CA for this ZED. If the color in the circle is white, it indicates that there is no position within the building where the BEP is less than 0.01 for that ZED, typically due to signal weakness. The influence of $N_{TTI}$ is visible on the CA of ZED 109 is depicted in grey-green, located at the lower-left region of the map. The CA significantly increases with the $N_{TTI}$. The CA strongly depends on the number of pilot symbols per bit N.

## VI. CONCLUSION

In this paper, we propose an ultra-low power new indoor localization scheme of smartphones that relies on zero-energy-devices beacons instead of active wireless beacons. The proposed scheme has been evaluated through extensive simulation corresponding to a real existing building covered by a real base station from Orange 4G commercial network. In these realistic conditions, we show that a room-level precision can be obtained with a basic localization scheme and a coherent detector. Next studies will assess more advanced localization schemes and non-coherent detectors.

## ACKNOWLEDGEMENT

This work is in part supported by the European Project Hexa-X II under (grant 101095759).

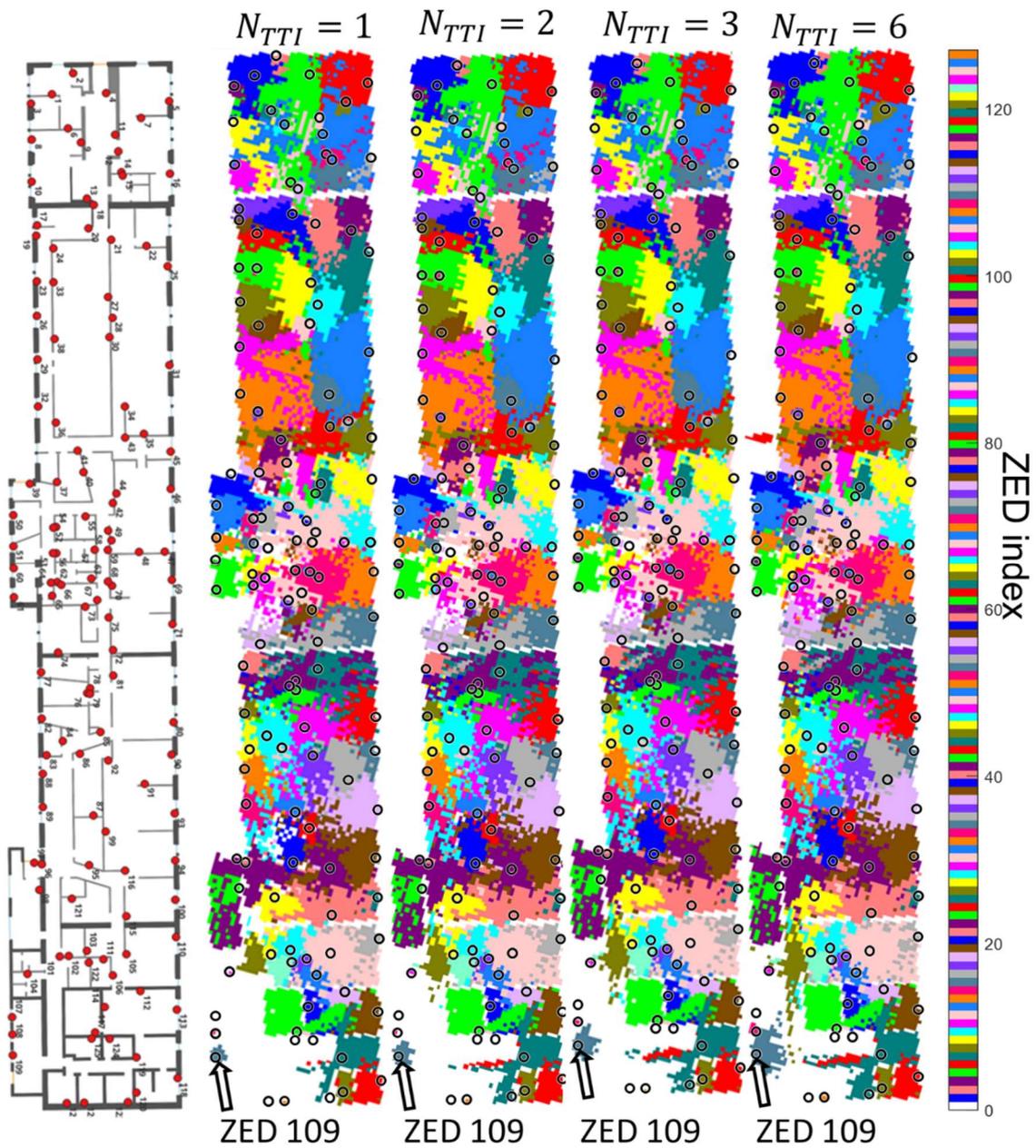

Fig. 5.  Map of deployment on the left and CA simulated with different values of $N_{TTI} = 1$, 2, 3 and 6